\documentclass[%
 reprint,
showpacs,
amsmath,amssymb,
aps,
prb,
]{revtex4-1}

\usepackage[usenames,dvipsnames,svgnames,table]{xcolor}
\usepackage{float}
\usepackage{graphicx}
\usepackage{dcolumn}
\usepackage{bm}

\usepackage[
margin=0.7in,
]{geometry}

\newcommand{\EqLabel}[1]{\label{#1}}

\begin{document}
 
\title{Phonon-assisted carrier motion on the Wannier-Stark ladder}

\author{Alfred Ka Chun Cheung}

\affiliation{Department of Physics and Astronomy, University of
  British Columbia, Vancouver, BC V6T 1Z1, Canada }

\author{Mona Berciu}

\affiliation{Department of Physics and Astronomy, University of
  British Columbia, Vancouver, BC V6T 1Z1, Canada }

\affiliation{Quantum Matter Institute, University of
  British Columbia, Vancouver, BC V6T 1Z4, Canada }

\date{\today}
\begin{abstract}
It is well known that at zero temperature and in the absence of
electron-phonon coupling, the presence of an electric field leads
to localization of carriers residing in a single  band of finite
bandwidth. We implement the Self-Consistent Born
Approximation (SCBA) to study the effect of weak electron-phonon
coupling on the motion of a carrier in a biased system. At moderate
and strong electron-phonon 
coupling we supplement the SCBA, describing the string of phonons
left behind by the carrier, with the Momentum Average (MA)
approximation to describe the phonon cloud that accompanies the
resulting polaron. We find that
coupling to the lattice delocalizes the carrier, as expected,
although long-lived resonances resulting from the Wannier-Stark states of
the polaron may appear in the spectrum in certain regions of the parameter
space. The approach we propose here can  also be used
to implement and check the validity of simple variational approximations. 
\end{abstract}

\maketitle

\section{Introduction}
 
 It has long been known\cite{nenciu,super} that carriers in a clean one-dimensional
 tight-binding band become localized \cite{gluck} if a uniform
 electric field $E$ is applied, since this breaks the free-particle
 continuum into a sequence of equally spaced discrete levels separated
 by the electric potential energy between consecutive sites
 $\delta = e a E$, where $e$ is the carrier's charge and $a$ the
 lattice constant. This discrete spectrum is the Wannier-Stark (WS)
 ladder.\cite{wannier}

\begin{figure}[b]
\includegraphics[width=\columnwidth]{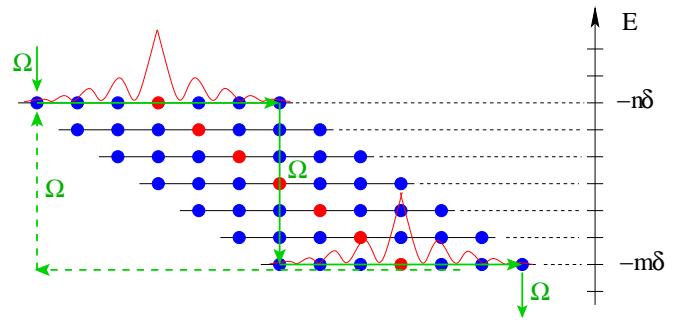}
\caption{(color online) Carrier motion on the WS ladder. The
  horizontal axis is the chain, with sites shown as dots. The vertical
  axis is the energy. Several WS eigenstates are shown, each centered
  at the site (red dot) with the same on-site energy. The probability
  distribution is sketched for two of these. Green arrows show part of
  the evolution, with the carrier arriving on the $-n\delta$ level
  from a higher one upon phonon emission; it then hops towards right
  and eventually emits another phonon to move on the $-m \delta$
  level, etc. If phonons are absorbed in reverse emission order when
  the carrier retraces its steps, then only non-crossed diagrams  are generated.  Crossed diagrams are for processes shown by the dashed line, where
  phonons are not absorbed in reverse emission order.
\label{fig1}}
\end{figure}

While the ladder has been observed in semiconductor superlattices and
in cold-atom systems,\cite{super} it is not seen in the spectra of
regular crystalline solids. The absence of localization is easily
understood in metals, because the Fermi sea electrons screen out the
electric field and carriers move ballistically as described by the
Buttiker-Landauer theory\cite{But} (if correlations can be neglected).
In insulators, however, the electric field is not screened out and
therefore the band is ``tilted''. Here, the absence of localization is
attributed to coupling to the lattice: a carrier can emit
phonons\cite{Emin} and thus lower its energy to slide along the chain,
as sketched in Fig. \ref{fig1}.  Most previous work on this problem
assumes incoherent tunneling between sites.\cite{Emin} For example,
this is routinely done when modeling carrier transport in organic
solar cells, based on the belief that those organic
semiconductors are so disordered as to destroy coherence.\cite{deibel}
While this assumption awaits validation, an understanding of the full
quantum dynamics, which should be relevant in clean(er)
systems,\cite{Mop} is still needed.
 
The quantum problem was first studied numerically in
Ref. \onlinecite{trug1}, with a variational solution assuming that
phonons appear only on the same site or to the left (uphill) of the
carrier. Ref.  \onlinecite{ulloa} obtained analytic and numerical
results for the spectrum of a finite chain for weak electron-phonon
(e-ph) coupling and small hopping $t \ll \delta$, while
Ref. \onlinecite{trug2} investigated the time evolution of the
wave-function once the electric field is turned on.

The method we propose here is similar in spirit to that used in 
Ref. \onlinecite{trug1}, however we use a different assumption to
calculate analytically the Green's function for this problem. Unlike
Ref. \onlinecite{trug1}, we do not restrict the direction of motion of
the carrier, instead we assume that the phonons left behind by the
carrier can only be absorbed in inverse order to that in which they
were emitted. This leads to only non-crossed diagrams being summed in
such processes,
which is the essence of the Self-Consistent Born Approximation
(SCBA). For the non-biased system ($\delta=0$), SCBA is known to be
accurate only at weak e-ph coupling. For moderate and strong e-ph
coupling, we use SCBA to describe this string of phonons left behind
as the carrier moves to lower energies, and combine it with the Momentum Average (MA)
approximation to describe the phonon cloud that accompanies
the carrier, turning it into a polaron. MA has been shown to provide a
rather accurate description of the polaron properties for any e-ph
coupling strength in an unbiased system, so long as the energy of the
optical phonons, $\Omega$, is not 
too small.\cite{MA} 

We argue that taken together, these approximations
allow us to understand the local density of states (LDOS) in such
a system if the bias $\delta$ is not large compared to 
$\Omega$. Our results uncover  the evolution of the LDOS as the e-ph
coupling is turned on, confirming that delocalization occurs as soon
as such coupling is present. However, for strong e-ph coupling and
smaller biases, very sharp resonances can appear in the spectrum, and
are understood as being due to WS-like states for the polaron,
which however can tunnel into extended states located further
downhill. We believe that these results supplement those presented in
Refs. \onlinecite{trug1,ulloa,trug2} to improve our
understanding of the quantum dynamics in this system. The formalism we propose here can also
be easily modified to implement  other variational descriptions
to check for their validity, as we exemplify for two particular
cases. Other possible generalizations are discussed at the end. 

The article is organized as follows: Section II describes the model
and the formalism we use to calculate the propagators and resulting
LDOS. The results are presented in Section III, while Section IV
contains a summary and some further discussions.

\section{Model and formalism } 

The model Hamiltonian we study is described by:
\begin{equation}
\EqLabel{1} {\cal H} = {\cal H}_{\rm e} + {\cal H}_{\rm ph} + V_{\rm
  e-ph}
\end{equation}
where $${\cal H}_{\rm e} = -t \sum_{n}^{} (c_n^\dagger c_{n+1} + h.c. )
+ \sum_{n}^{} \epsilon_n c_n^\dagger c_n$$ describes nearest-neighbor
hopping of the carrier on a 1D chain biased by the
applied electric field, so that the on-site energies are $\epsilon_n = -neaE=
- n \delta$. (The spin is trivial and we ignore it for
simplicity). There is an Einstein phonon mode $$
{\cal H}_{\rm ph} = \Omega \sum_{n}^{} b_n^\dagger b_n$$
(for simplicity, we take $\hbar =1$ in the following). 
Finally, $$V_{\rm
  e-ph} = g \sum_{n}^{} c^\dagger_n c_n ( b^\dagger_n +b_n)$$ is the
Holstein model\cite{Holstein} for e-ph coupling. 
As usual, $c_n$ and $ b_n$ are annihilation
operators for the carrier and phonons, respectively, at site $n$ of
the chain. Also as customary, we will gauge the strength of the e-ph
coupling with the dimensionless effective coupling:
$$
\lambda = {g^2\over 2t\Omega}
$$
 appropriate for 1D models.\cite{MA}

The quantity of interest is $G(n,z) = \langle 0 | c_0 \hat{G}(z)
c^\dagger_n|0\rangle$ where $|0\rangle$ is the vacuum  and
$\hat{G}(z) = [z - {\cal H}]^{-1}$ is the resolvent at
$z=\omega + i \eta$, where $\eta \rightarrow 0$ controls the artificial
lifetime $\sim 1/\eta$ of the carrier. This is the Fourier
transform of  $G(n,\tau) \sim \Theta(\tau)
\langle 0 | c_0 e^{-i {\cal H} \tau} c^\dagger_n|0\rangle$, {\em i.e.}
the amplitude of probability for the carrier to move from site $n$
to site $0$ in a time $\tau$ and so that all phonons emitted in the
meantime have been re-absorbed. If such a process is very unlikely, then
$G(n,z) \rightarrow 0$. From the Lehmann representation\cite{mahan} we
know that 
the local density of states (LDOS) $A(n,\omega) = -{1\over \pi}
\mbox{Im} G(n,z)$ is finite at energies $\omega=E_\alpha$ in the
one-carrier spectrum ${\cal H} |\phi_\alpha\rangle = E_\alpha
|\phi_\alpha\rangle$, provided that the overlaps $\langle 0| c_0
|\phi_\alpha\rangle \langle \phi_\alpha| c_n^\dagger|0\rangle$ do not
vanish. As will become apparent soon, our method to calculate $G(n,z)$
also gives the generalized propagators:
\begin{equation}
\EqLabel{2} F_k(n; n_k, \dots,n_1; z) = \langle 0 | c_0 \hat{G}(z)
c^\dagger_nb_{n_k}^\dagger \cdots b_{n_1}^\dagger|0\rangle
\end{equation}
whose meaning and usefulness mirror those of $G(n,z)$.

\subsection{No e-ph coupling: $\lambda=0$}

We first 
calculate  $G_0(n,z)=\langle 0 | c_0 \hat{G}_{\rm e}(z)
c^\dagger_n|0\rangle$ for ${\cal H}_e$, {\em i.e.} 
when there is no e-ph coupling. Taking appropriate matrix
elements of the identity $\hat{G}_{\rm e}(z)(z-{\cal H}_{\rm e} )  =1$
gives the equations of motion (EOM):
\begin{equation}
\EqLabel{3}
(z-\epsilon_n) G_0(n,z) = \delta_{n,0} -t[G_0(n-1,z)+ G_0(n+1,z)]
\end{equation}
These are solved easily if we recognize that $G_0(n,z) \rightarrow
0 $ for sufficiently large $|n|$ because the electron cannot move
arbitrarily far in a finite lifetime $1/\eta$. As a result,
\begin{equation}
\EqLabel{4}
\begin{array}[c]{cc}
G_0(n,z) = A(z-\epsilon_n) G_0(n-1,z) & \mbox{  , if  } n> 0\\
G_0(n,z) = B(z-\epsilon_n) G_0(n+1,z) & \mbox{  , if  } n< 0 \\
\end{array}
\end{equation}
where  we define the continued fractions:
\begin{equation}
\EqLabel{4b}
\begin{array}[c]{c}
A(f(z)) = \cfrac{-t}{f(z) +t A(f(z+\delta))} \\
B(f(z)) = \cfrac{-t}{f(z) +t B(f(z-\delta))}.\\
\end{array}
\end{equation}
These quantities 
are calculated
iterationally starting from a cutoff $A(z+N \delta)=B(z-N\delta)=0$  for a
sufficiently large $N$. Because for $\delta \ne 0$
all eigenstates are localized, a
cutoff $N\sim 20$ usually suffices. If the electric field is
turned off, $\delta=0$, they can be found
analytically to be $A(z)|_{\delta=0} =-B(-z)|_{\delta=0} = -z/ 2t + \sqrt{z/
2t +1} \sqrt{z/  2t -1}$ so that $|A(z)| <1$, $|B(z)| <1$,   if
$\mbox{Im}(z)=\eta >0$. Finally, using the $n=\pm 1$ results of Eq. (\ref{4}) in
Eq. (\ref{3}) leads to:
\begin{equation}
\EqLabel{5}
\begin{array}[c]{cc}
G_0(0,z) = \cfrac{1}{ z + t[A(z+\delta)+B(z-\delta)]}& , n=0\\
G_0(n,z) = A(z+n\delta)\cdots A(z+\delta) G_0(0,z) &,  n
>0\\
G_0(n,z) = B(z-n\delta) \cdots B(z-\delta) G_0(0,z) & , n<0. \nonumber \\
\end{array}
\end{equation}
If $\delta=0$  this gives the usual
results for a tight-binding model.\cite{Economou} For
 $\delta\ne 0$  it is easy to check that
the WS energies $E_n=n\delta$ are indeed poles of
$G_0(n,z)$. Full mapping onto 
the analytic solution\cite{Fukuyama} can also be
verified.\cite{davison} 

\subsection{Weak e-ph coupling, $\lambda \ll 1$: SCBA}

For $g\ne 0$, the EOM acquire additional terms because of phonon
emission and absorption. In particular, now:
\begin{multline}
\EqLabel{6}
(z-\epsilon_n) G(n,z) = \delta_{n,0} -t[G(n-1,z)+ G(n+1,z)]\\ 
+gF_1(n;n;\omega)
\end{multline}
Exact EOM for  $F_k, k\ge 1$ of Eq. (\ref{2}) can
be easily derived, however the resulting infinite system of coupled
equations is too complicated, thus approximations are needed.

Physically, we expect the carrier to leave phonons behind, as sketched
in Fig. \ref{fig1}, in order to move down the ladder.  The more
probable processes, leading to diagrams with the largest
contributions, are like those shown by full lines: phonons are emitted
when needed to move between different ladder states and are absorbed
in reverse order if the carrier goes back. A process leading to a
crossed diagram is shown by the dashed lines, and should have a low
probability because the ladder states are localized. Note that here we
assume that the phonons left behind are typically not spatially very close to
one another. This is a reasonable
assumption  if $\Omega > \delta$. Below, we
will also gauge the validity of this assumption in the case where $\Omega
\sim \delta$.

The assumption that the contribution of crossed diagrams can be ignored
is the essence of the Self-Consistent Born approximation (SCBA). For
weak coupling $\lambda \ll1$, SCBA is known to be a reasonable
approximation in the un-biased system with $\delta=0$.\cite{MA} This
is another reason to expect that its generalization to the biased case, provided
here, should continue to work well for small $\lambda$.

By keeping only non-crossed diagrams, SCBA assumes that phonons are
absorbed in inverse order to their emission order, {\em i.e.}  if
phonons were priorly emitted (in this order) at sites $n_1, \dots,
n_k$, at this point either another phonon is emitted, or only the one
at $n_k$ can be absorbed. For this to be possible, these phonons must
be distinguishable. This is automatically the case if they are  located
at different sites. If there are multiple phonons emitted at the same
site, we will treat them as if they belong to different phonon modes
so that they continue to be distinguishable. As we show below,
this is implicitly assumed to be true for SCBA in the unbiased system
with $\delta=0$. It should remain a reasonable assumption for the biased case as
well if $\Omega > \delta$, since, as already discussed, we do not
expect multiple phonons to be located at the same site with high
probability. Below we provide a 
way to gauge the validity of this approximation.

After imposing these restrictions, the EOM for the generalized
propagator $F_k(n; n_k, \dots, n_1; 
z)$, $k\ge 1$, read:
\begin{widetext}
\begin{multline}
\EqLabel{7}
(z-\epsilon_{n_k}-k\Omega) F_k(n_k; n_k, \dots, n_1; z) = - t
  \left[F_k(n_k-1; n_k, \dots, n_1; z)+F_k(n_k+1; n_k, \dots, n_1;
    z)\right] \\ + g F_{k-1}(n_k; n_{k-1} \dots, n_1; z)
+ g F_{k+1}(n_k; n_k, n_k, n_{k-1} \dots, n_1; z)
\end{multline}
if $n=n_k$, while for $n\ne n_k$:
\begin{equation}
\EqLabel{8}
(z-\epsilon_{n}-k\Omega) F_k(n; n_k, \dots, n_1; z) = - t
  \left[F_k(n-1; n_k, \dots, n_1; z)+F_k(n+1; n_k, \dots, n_1;
    z)\right] + g F_{k+1}(n;n, n_k, \dots, n_1; z).
\end{equation}
In other words, if the carrier is at the site $n=n_k$ where the last
emitted phonon resides, it can  hop away, it can absorb that phonon
or it can create another phonon (treated as if it belongs to a
different mode) at the same site. If the carrier is at
a site $n\ne n_k$  it can  hop away or emit a new
phonon, but absorption of one of the existing phonons is
not allowed by the non-crossing condition.

Remarkably, these EOM can be
solved analytically by noting that for any $k\ge 0$  we must have 
\begin{equation}
\EqLabel{9}
 F_{k+1}(n; n,n_k, \dots, n_1; z) = \sigma(z-\epsilon_n - k\Omega)
 F_{k}(n; n_k, \dots, n_1; z).
\end{equation}
Mathematically, this is because if $k$ and $n$ are large enough, these
 propagators must eventually vanish. Truncating the EOM  at any
 $k+2$  ($k$ can be 
 arbitrarily large) leads to a form similar to Eq. (\ref{9}). This
 ansatz turns Eq. (\ref{8}) into a simple 
 recurrence equation like Eq. (\ref{3}), thus $F_k(n_k+1;
 n_k,\dots,n_1;z) = A(z-\epsilon_{n_k+1} -k\Omega -
 g\sigma(z-\epsilon_{n_k+1} -k\Omega)) F_k(n_k, n_k,\dots,n_1;z)$,
 and $F_k(n_k-1;
 n_k,\dots,n_1;z) = B(z-\epsilon_{n_k-1} -k\Omega -
 g\sigma(z-\epsilon_{n_k-1} -k\Omega)) F_k(n_k, n_k,\dots,n_1;z)$. Using these in
 Eq. (\ref{7}) leads to an equation consistent with the ansatz of
 Eq. (\ref{9}), from which 
 we find:
\begin{equation}
\EqLabel{10}
\sigma(z) = \frac{g}{z-\Omega -g\sigma(z-\Omega) +
  tA(z+\delta-\Omega-g\sigma(z+\delta-\Omega)) + tB
  (z-\delta-\Omega-g\sigma(z-\delta-\Omega)) }.
\end{equation}
\end{widetext}
The solution of this equation   can be calculated iterationally starting from
$\sigma(z) \approx 1/(z-\Omega) $ as $|z|\rightarrow \infty$.

Physically, Eq. (\ref{9}) means that the amplitude of probability for
an additional phonon to be emitted depends only on the energy of the
electron, and not on the detailed locations of the previously emitted
phonons.  Using the ansatz for $k=1$ into Eq. (\ref{6}) leads to:
\begin{equation}
\EqLabel{10b}
G(n,\omega) = G_0(n,\omega- g\sigma(\omega))
\end{equation}
and we recognize $\Sigma_{\rm SCBA}(\omega) = g \sigma(\omega)$. It is
straightforward to verify that for $\delta=0$, this is the expected
solution $\Sigma_{\rm SCBA}(\omega)={g^2\over N}
\sum_{q}^{}G_{\rm SCBA}(k-q,\omega-\Omega)$ where $ G_{\rm
  SCBA}(k,\omega)=1/\left[\omega+i \eta - \epsilon_k -\Sigma_{\rm
    SCBA}(\omega)\right] $.\cite{MA} One can now 
obtain the SCBA values for  other propagators $F_k$.

To check the validity of this approximation, we can use the same
framework to implement other variational schemes and compare the
results.  For example, the solution of Ref. \onlinecite{trug1} can be
trivially implemented by setting in the EOM $F_k(n; n_k, \dots,
n_1,\omega)=0$ if $n <n_k$, i.e. the carrier cannot be to the left of
the last emitted phonon (this automatically implies $n_1 \le \dots \le
n_k$). As a result, the corresponding self-energy (which we label as
``Ref. [9]'' in the following) has $B \equiv 0$ in Eq. (\ref{10}).  A
priori, we do not expect this approximation to be that good for very small
biases where the effective probabilities for the carrier to hop uphill
vs. downhill are not that different.

A wider variational space can be achieved by allowing the electron to
go anywhere but keeping the additional restriction $n_1 \le \dots \le
n_k$, i.e. the electron can move to the left of existing phonons but
it cannot emit additional phonons while there. Since this is one of
the ways to obtain multiple phonons at the same site, this
approximation allows us to gauge the importance of contributions from
configurations with multiple phonons at the same site. Mathematically,
the corresponding EOM for this variational approximation (which we
label ``var'' in the following) are obtained by removing the last term
in Eq. (\ref{8}) when $n < n_k$.  Its solution is like Eq. (\ref{10})
but with $B(z-\delta -\Omega)$ in the denominator (since no new
phonons are emitted when the carrier moves to the left of existing
phonons, the contribution from such paths is not renormalized by the
self-energy $g\sigma$).  Various other possibilities can be
implemented similarly, by only keeping terms in the EOM consistent
with those assumptions, but we stop here.

\subsection{Moderate and large e-ph coupling: MA+SCBA}

For stronger electron-phonon coupling, the probability to have
multiple phonons at the same site must increase. This is known to be
the case even for the un-biased system, because the electron creates a
robust phonon cloud that accompanies it as it moves through the
system. The resulting dressed quasiparticle is, of course, the
polaron. In the biased system, one would expect the polaron to move
down the ladder with its robust cloud. 

In such conditions, we expect that the approximation made above, of
treating multiple phonons that happen to be at the same site as if
they belong to different modes, to become quantitatively inaccurate
because of normalization factors. To see why, consider a state with
$n$ bosons at the same site. If they belong to the same mode, it is
described by $|n\rangle=b^{\dagger, n}/\sqrt{n}|0\rangle$, and we have
$b|n\rangle = \sqrt{n}|n-1\rangle$, etc. However, if we treat the $n$
bosons as belonging to $n$ distinct modes with one boson each, then
there are no normalization factors. For small $n\sim 1$ this makes little
difference, but this is no longer the case if many bosons are likely to
occur at the same site. 

Thus, at moderate and large $\lambda$, the e-ph coupling has two
consequences: one is to lead to the formation of the polaron with its
robust cloud, and the other is to allow it to move to lower energies
by leaving phonons behind. The number and typical locations of the
phonons left behind is controlled by the ratio $\Omega/\delta$ and
therefore is not very sensitive to the strength of the coupling. As a
result, we expect these processes to continue to be well described by
the SCBA scheme, i.e. by assuming that these
phonons are only involved in non-crossed diagrams.

However, at any point the electron can start building a larger cloud
in its vicinity (the polaron cloud). Since this cloud typically
contains many phonons, it is unlikely that the electron will abandon
it and move away to start building another robust cloud, at least not
in the case $\Omega \gtrsim \delta$ that we consider here. Instead,
the electron will reabsorb these cloud phonons and then move to
another location
(maybe leaving one phonon behind) and start creating another robust polaron
cloud, similarly to how it moves in an unbiased system. 

In the unbiased system and for moderate and large e-ph coupling, it
has been shown that the Momentum Average (MA) approximation\cite{MA}
provides an accurate description of the polaron properties so long as
$\Omega$ is not very small. For the Holstein model, MA has been shown
to correspond to the variational approximation of assuming that the
polaron cloud has all its phonons at one site.\cite{var} This
variational space can be enlarged systematically to check its
validity. While for the Holstein model this approximation is already
very reasonable, for more complicated models of e-ph coupling one
needs to allow the polaron cloud to spread over multiple adjacent
sites.\cite{other}

Here we implement an MA+SCBA approximation which  assumes that a one-site
polaron cloud can only be built at the location of the last emitted
phonon, and that while a cloud with two or more phonons is
present the electron will not emit/absorb phonons in other
locations, consistent with MA. At the same time, phonons not in the cloud can only be
absorbed in the inverse order to that in which they were emitted, as
described by SCBA.

Mathematically, we implement this as follows. The EOM
remain unchanged as long as $n_k \ne n_{k-1}$, i.e. no cloud is being
built. Equations (\ref{8}) and (\ref{9})  are supplemented with  additional equations
for the propagators with multiple phonons at site $n_k$. Specifically,
for any $p\ge 1$, and using the shorthand notation $\{ n\}_{p+1} \equiv n_k,
\dots, n_k, n_{k-1}, \dots , n_1$ where the first $p+1$ sites are all $n_k$, we have:\cite{comment}
\begin{widetext}
\begin{multline}
\EqLabel{7n}
(z-\epsilon_{n_k}-(k+p)\Omega) F_{k+p}(n_k; \{n\}_{p+1}; z) = - t 
  \left[F_{k+p}(n_k-1; \{n\}_{p+1}; z)+F_{k+p}(n_k+1;  \{n\}_{p+1};
    z)\right] \\ + (p+1) g F_{k+p-1}(n_k;  \{n\}_{p}; z)
+ g F_{k+p+1}(n_k;  \{n\}_{p+2}; z)
\end{multline}
for  $n=n_k$, while for $n\ne n_k$:
\begin{equation}
\EqLabel{8n}
(z-\epsilon_{n}-(k+p)\Omega) F_{k+p}(n; \{n\}_{p+1}; z) = - t 
  \left[F_{k+p}(n-1; \{n\}_{p+1}; z)+F_{k+p}(n+1; \{n\}_{p+1};
    z)\right] 
\end{equation}
These additional equations can be solved trivially and give:
\begin{equation}
\EqLabel{9n}
F_{k+1}(n_k; n_k, n_k, n_{k-1},\dots, n_1; z) =
\sigma_{MA}(z-\epsilon_{n_k}-(k+1)\Omega) F_{k}(n_k; n_k, \dots, n_1; z)
\end{equation}
where
\begin{equation}
\EqLabel{MA}
\sigma_{MA}(z) = \cfrac{2g}{z + tA(z+\delta)+tB(z-\delta)-
  \cfrac{3g^2}{z -\Omega +
    tA(z-\Omega+\delta)+tB(z-\Omega-\delta)-\cfrac{4g^2}{\dots}}} 
\end{equation}
The ansatz of Eq. (\ref{9}) remains unchanged if the last two phonons
are not at the same site, and is supplemented by Eq. (\ref{9n}) if the
last two phonons are at the same site. The rest of the
solution proceeds as before and we find
\begin{equation}
\EqLabel{16}
\sigma(z) = \frac{g}{z-\Omega -g\sigma_{MA}(z-\Omega) +
  tA(z+\delta-\Omega-g\sigma(z+\delta-\Omega)) + tB
  (z-\delta-\Omega-g\sigma(z-\delta-\Omega)) }.
\end{equation}
\end{widetext}
 
Again, we will check this approximation against the
variational predictions that do not allow the electron to move to the
left of the rightmost phonon (labeled as ``Ref. [9]''), respectively
allow it to do so but not to emit additional phonons to the left of
the rightmost one (labeled as ``var''). These are implemented just as
before. Another approximation we implement, which will be labeled as
``MA+SCBA dressed'', is obtained by replacing $A(z) \rightarrow
A(z-g\sigma(z)), B(z) \rightarrow B(z-g\sigma(z))$ everywhere in
Eq. (\ref{MA}). As its name suggests, this approximation allows the
electron to start building additional non-crossed strings of phonons, which may
include one-site larger clouds, while the original cloud is
present, because it is obtained by replacing bare
propagators by full propagators in Eq. (\ref{MA}). Comparing it to the
MA+SCBA results will allow us to gauge whether the assumption that
such processes can be ignored is correct.

\section{Results}

\subsection{Weak coupling limit: SCBA}

\begin{figure}[t]
\includegraphics[width=0.8\columnwidth]{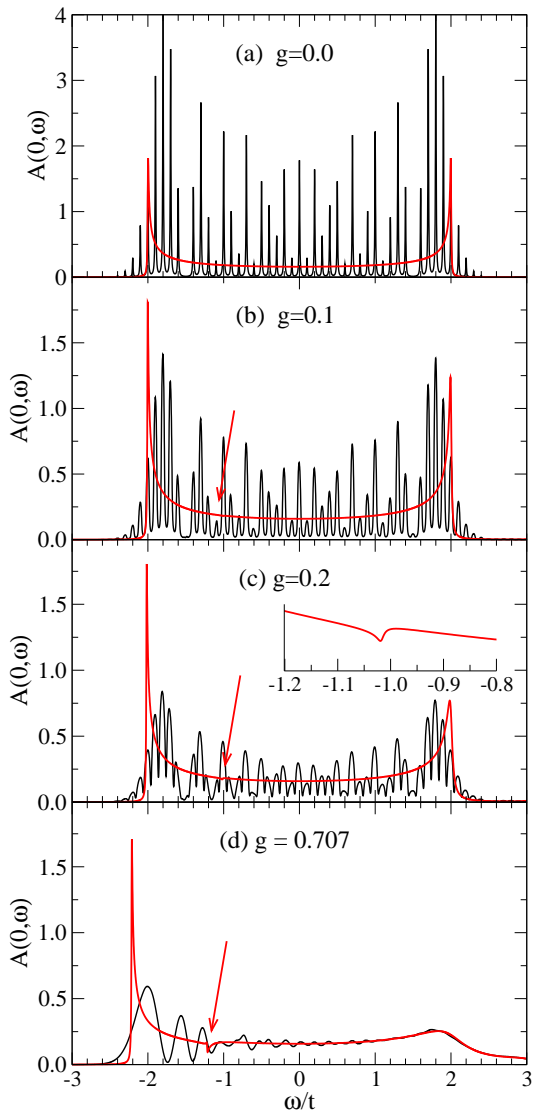}
\caption{(color online) $A(0,\omega)$ vs $\omega$ for
 $\lambda=0, 0.005, 0.02$ and 0.25 and $t=1, \Omega=1,
  \eta=0.005$. The thin black line shows results for 
  $\delta=0.1\Omega$ while the thick red line is for $\delta=0$. Arrows mark the top
  of the polaron band, also see inset.
\label{fig2}}
\end{figure}

We begin by analyzing a system with a small bias  and small e-ph
coupling, using the SCBA approximation. Typical results are shown in
Fig. \ref{fig2}, which plots the $n=0$ LDOS for various values of $g$,  
with (thin black line) and without (thick red line) an electric field
$\delta=0.1$ for $\Omega=t=1$. The LDOS at other sites is given by
$A(n,\omega) = A(0, \omega-\epsilon_n)$, i.e. it is shifted by $n
\delta$. 

Figure \ref{fig2}(a) shows the $g=0$ results. As expected, the biased
system's LDOS shows discrete peaks at $\omega=m\delta$ marking the
WS ladder. Some of these peaks are hard to see because their
wave-function is very small at 
site $n=0$. This is progressively the case for peaks with energies
$|\omega| > 2.5t$ because of their localized nature. The LDOS of the
unbiased system is the usual 1D result, with a continuum of states for
$|\omega|\le 2t$.

As we turn the e-ph coupling on in panels (b)-(d), the former WS
states acquire a finite lifetime (their width is no longer controlled
by $\eta$, instead being significantly larger even for $g=0.1$, see
change in the vertical scale), showing that these states are no longer
localized. This proves that coupling to the lattice indeed results in
delocalization. As $\lambda$ increases, the peaks continue to broaden
and start to merge into a smooth continuum. This occurs in an
asymmetric way, with higher energy states converging faster towards a
smooth LDOS, while the lower energy states still show considerable
LDOS variation.

This may seem surprising at first, but the reason becomes clear when
we compare with the LDOS for $\delta=0$ (thick red line), which has
two features: a polaron band at low energies $\omega \in [E_{GS},
  E_{GS}+\Omega]$ ($E_{GS}$ is the polaron ground-state energy) and
the polaron+one-phonon continuum for $\omega >
E_{GS}+\Omega$.\cite{var} Arrows mark the transition between the two features,
which is barely visible on this scale for $g=0.1, 0.2$ (for the later
case, it is shown more clearly in the inset). States in the polaron
band describe the coherent, infinite-lifetime quasiparticle (the
polaron) consisting of the carrier and its phonon cloud. In contrast,
the polaron+one-phonon continuum contains incoherent states with
finite lifetime, describing the scattering of the polaron on one or
more phonons that do not belong to its cloud.

At first, one may expect that turning on an electric field should have
a very different effect on the two types of states: the incoherent
states at high energy should remain delocalized since the polaron
already has enough energy to leave phonons behind and can continue to
do so when the bias is applied. However, at low
energies one may expect to see a WS ladder describing the localization
of the polaron. Indeed, if here the polaron carries all the phonons in its
cloud then it cannot leave any of them behind, therefore the electric
field should localize it just like it does with a bare particle. However, because
the LDOS at site $n$ is shifted downward by $n\delta$, it follows that
such localized WS states could tunnel into the continuum that appears
at the same energies for sufficiently large $n>0$. In other words,
such states cannot be localized, instead they are at most resonances with a width
controlled by the tunneling rate. If this is large compared to
$\delta$ then the resonances merge into a smoother LDOS, as we see
for these parameters. Indeed, as shown below, individual resonances
spaced by $\delta$ can be recovered either by increasing $\delta$,
and/or by increasing the e-ph coupling, which makes the polaron very
heavy and therefore greatly decreases its tunneling rate.

\begin{figure}[t]
\includegraphics[width=0.8\columnwidth]{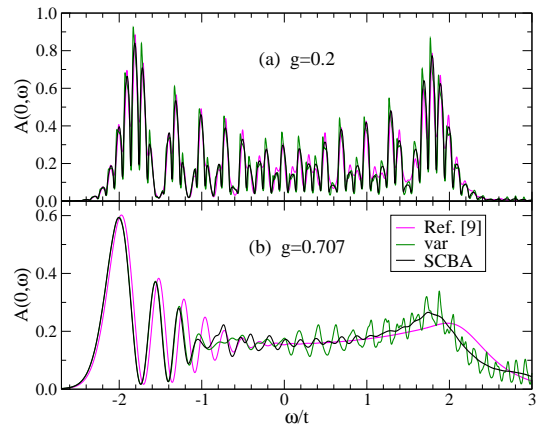}
\caption{(color online) Comparison between the SCBA LDOS and those
  predicted by  the ``Ref. \onlinecite{trug1}'' and ``var''
  approximations (see text for 
  more details), for parameters as in Fig. \ref{fig2}.
\label{fig3}}
\end{figure}

Before looking at other parameters, we compare the results of
SCBA with those of the other two variational approximations
discussed. This comparison is shown in Fig. \ref{fig3} for two values
of $g$. For the lower value we see very good agreement between all
three curves, confirming that here it is indeed very unlikely for the
electron to return past emitted phonons. As $g$ increases, however,
the approximation of Ref. \onlinecite{trug1} becomes less accurate,
while the variational approximation which allows the
electron to move to the left of the existing phonons but not to emit
other phonons there, is still extremely accurate at low energies (here
the two curves are indistinguishable). This shows
that it is not likely for the electron to return and emit more
phonons to sites where it already emitted phonons in the past,
validating our assumption that sites with multiple phonons are very
few for these parameters.

\begin{figure}[t]
\includegraphics[width=0.8\columnwidth]{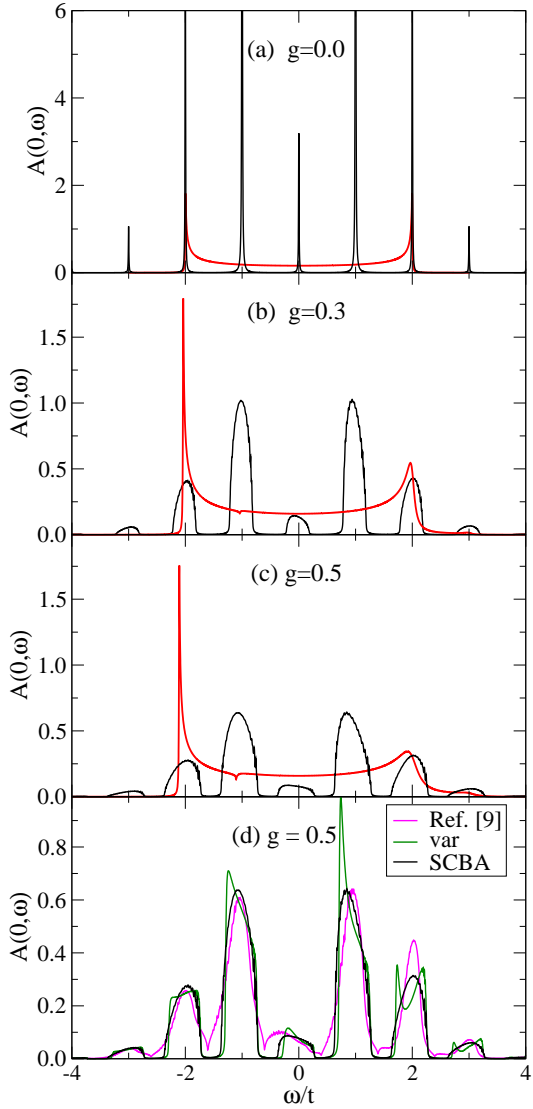}
\caption{(color online) Panels (a)-(c) show $A(0,\omega)$ vs $\omega$ for
 $\lambda=0, 0.045$ and 0.125 and $t=1, \Omega=1,
  \eta=0.005$. The thin black line shows results for 
  $\delta=\Omega$ while the thick red line is for $\delta=0$. Panel
  (d) compares the three approximations (see text for details).
\label{fig4}}
\end{figure}

In Fig. \ref{fig4} we show results for similarly small e-ph couplings
but a much larger bias $\delta = \Omega$. Here, the broadening of the
former WS states into resonances as the e-ph is turned on is very
clearly visible, with their width increasing with $\lambda$. Because
$\delta$ is so large, these resonances have not yet merged into a
continuum even at higher energies (this occurs at larger e-ph
coupling, as shown below, but larger $\lambda$ is not reliably described by SCBA).
The comparison with the other two variational approximations, shown in
panel (d), again confirms better agreement with the assumption that
the electron is free to move everywhere so long as it does not emit
more phonons to the left of the last emitted one.

\begin{figure}[t]
\includegraphics[width=0.8\columnwidth]{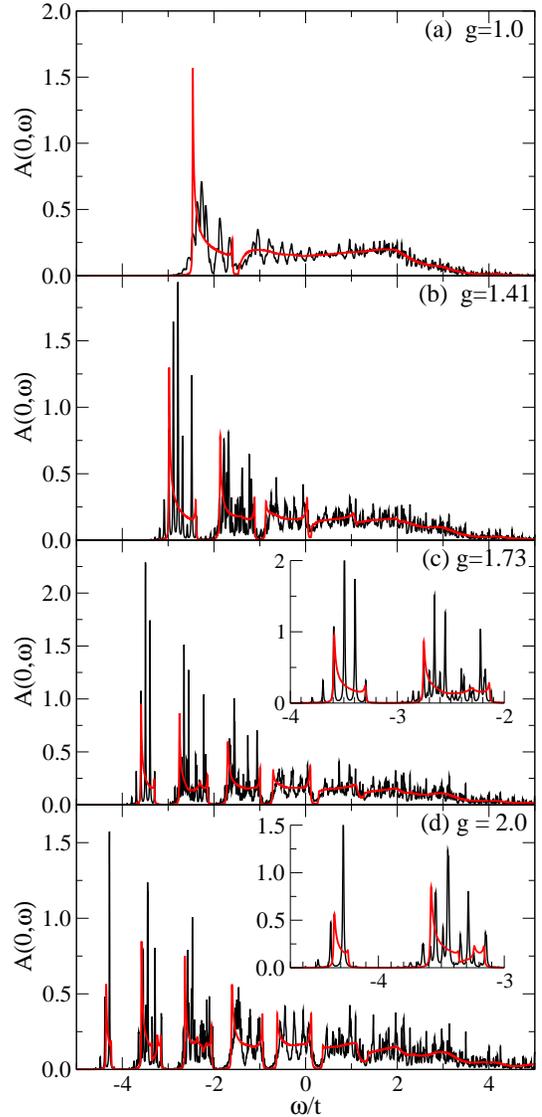}
\caption{(color online) LDOS for the small bias $\delta=0.1\Omega$
  when $t=\Omega=1$, but much larger e-ph couplings $\lambda=0.5, 1,
  1.5,2$. The thick red lines show the LDOS for $\delta=0$. The insets
  zoom into the low-energy sectors. 
\label{fig5}}
\end{figure}

The results in panels (b) and (c) should be compared with the two
lower curves in Fig. 4 of Ref. \onlinecite{trug1}, which plot the
current (not the LDOS) vs $\omega$ in a smaller range $\omega \in [-0.7, 1.7]$, and
also show gaps around $\omega=\pm 0.5, 1.5$ that decrease with
increasing $\lambda$. Their gaps are
smaller, and in fact are nearly closed for $g=0.5$ in agreement with
the results of panel (d), which compares the three
approximations. Panel (d) suggests that the variational approximation of
Ref. \onlinecite{trug1} overestimates the tunneling rate resulting in
broader peaks, although we must note that unlike SCBA, in 
Ref. \onlinecite{trug1} sites with multiple phonons are treated 
with the proper normalization factors. Despite these fairly minor quantitative
differences, however, it is clear that 
qualitatively all three approximations describe similar behavior, 
increasing our confidence that the exact solution is not too different.

\subsection{Moderate and strong coupling: MA+SCBA}

We now turn on the e-ph coupling and use the MA+SCBA method to study
the results (for the weak couplings discussed previously, there is no
difference between the MA+SCBA vs the SCBA results, as expected since
at weak couplings no robust phonon cloud forms).

In Fig. \ref{fig5}, we show results for the small bias $\delta=0.1\Omega$
but much larger $\lambda$ values. Consider first the results in the
un-biased case (thick red lines), which now show the polaron band
moving towards lower energies and becoming narrower, as $\lambda$
increases, as expected since the polaron becomes more stable but heavier. In panels
(c) and (d), for $\lambda=1.5$ and 2 respectively, the band associated
with the second bound state\cite{bonc} is also visible below the
continuum.\cite{MA}

\begin{figure}[t]
\includegraphics[width=0.8\columnwidth]{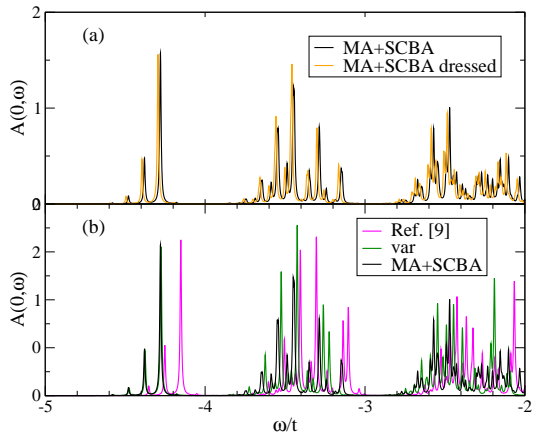}
\caption{(color online) Comparison between MA+SCBA and the various
  other approximations described in the text, for $\lambda=2$ and
  $10\delta=\Omega=t=1$. 
\label{fig6}}
\end{figure}

For a finite bias, the MA+SCBA results confirm our expectations
discussed above, namely that for heavier polarons the tunneling rates
are significantly decreased since moving towards right to tunnel into
the continuum becomes a very slow and therefore much less likely
process. Indeed, for the larger $\lambda$ 
values these tunneling rates are so small that the spectrum (at
energies corresponding to the polaron band) looks like a WS ladder
with the proper spacing $\delta$ between resonances, as seen more
clearly in the insets. At higher
energies the LDOS mimics the unbiased LDOS somewhat better, although
it still has significant ``peaky'' structure due to tunneling out of
the resonances lying further uphill.

In Fig. \ref{fig6} we compare the MA+SCBA results for $\lambda=2$ with the other
approximations described above. In particular, in panel (a) we compare
the low-energy sector of the $n=0$ LDOS to that predicted by the
dressed MA+SCBA approximation. The two curves are very similar apart
from a tiny shift due to the further renormalization of the polaron
cloud allowed by the dressed approximation, which  lowers its
energy. However, it is clear that this is a very small effect,
validating the assumption that a description of the phonon
configuration in terms of a one-site polaron cloud plus a  string of
phonons left behind  so that the polaron can lower its energy is
reasonable. Panel (b) shows the predictions of the other two
approximations, in very good agreement with MA+SCBA, at least at lower
energies. This is not surprising since while the robust polaron
cloud is present the electron is not expected to spend much time away
from the cloud site, therefore  additional restrictions on its
motion should indeed have little consequences.

\begin{figure}[t]
\includegraphics[width=0.8\columnwidth]{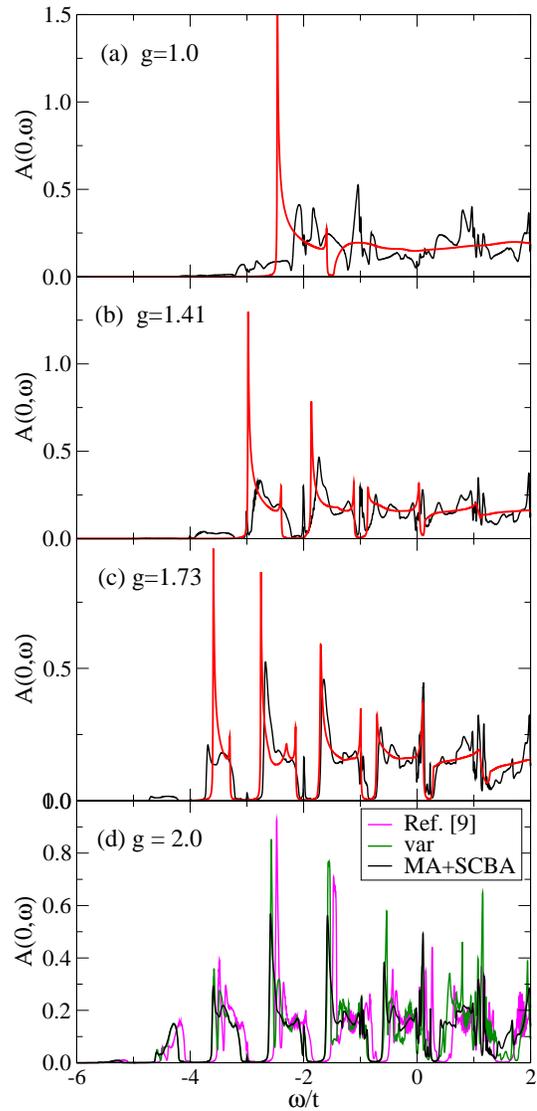}
\caption{(color online) LDOS for the large bias $\delta=\Omega$
  when $t=\Omega=1$, but much larger e-ph couplings $\lambda=0.5, 1,
  1.5,2$. The thick red lines in panels (a)-(c) show the LDOS for
  $\delta=0$. Panel (d) also shows the predictions of the other
  approximations discussed in the text.
\label{fig7}}
\end{figure}

Finally, in Fig. \ref{fig7} we show results for cases with large bias
$\delta=\Omega$ and strong coupling of up to $\lambda=2$. Individual
resonances associated with different WS-like states again become
visible at larger $\lambda$ (in particular, see feature appearing at
$\Omega$ below the polaron band) but are much broader than for the
small bias. This agrees with the trends observed at weak couplings,
and is expected since a larger bias must lead to increased tunneling
rates even for these heavy polarons.

Comparison between the different approximations, displayed in panel (d),
again shows good agreement. This suggests that the
assumption implemented in SCBA to describe the phonons left behind, as
the polaron moves further downhill, is still reasonable for a bias
$\delta \sim \Omega$. In other
words, a phonon is left behind every few sites, with low probability for
multiple phonons left at the same site or for phonons emitted later to be
to the left of phonons emitted earlier. For significantly larger bias
$\delta$ one expects this assumption to start to fail, since in this
case the carrier will need to emit many phonons at each site in order
to lower its energy enough to be able to delocalize effectively. As a
result, such cases cannot be described accurately by the approximations
we presented here.

\section{Summary and discussions} 

To summarize, we have implemented the SCBA approximation to describe
the string of phonons left behind by a carrier in a biased system, in
order to lower its energy to become delocalized. We argued that SCBA
should be provide a reasonable description for these processes  if the
bias is not too 
large. Increased coupling, however, also results in the dressing of
the carrier by a phonon cloud that accompanies it as it moves through
the system. Here we used the simplest variational MA flavor to describe
this cloud, combining it with SCBA to describe the phonons
left behind. We also showed how this formalism can be modified to
implement various other variational guesses that one might want to
test, and used two possible versions to validate our hypotheses for
certain parameter ranges.

Our results allow us to study the evolution of the spectrum as the
bias and/or the e-ph couplings are turned on. In is worth noting that
this Hamiltonian is rather unusual in that it has an unbounded
spectrum if the chain is infinite: moving further along the chain will
lower the energy arbitrarily much. However, we can calculate the LDOS
and use it to understand the states available in the vicinity of one
site. This can then be combined with the knowledge that at other sites
the LDOS looks similar, apart from the appropriate energy shift, to
gain a global understanding of its evolution.

We find that e-ph coupling always delocalizes the carrier, although
for large coupling and small biases one can observe sharp peaks in the
spectrum, that may be mistaken for localized states. As we argue, they
are in fact resonances because of tunneling into delocalized states
available further downhill.

While this method has been used here to study a clean 1D chain, both
SCBA and MA can be straightforwardly generalized to higher dimensions,
allowing this formalism to be used to investigate problems that become
progressively more difficult to study by numerical means.\cite{MA}
Other types of e-ph coupling can also be studied by similar
means,\cite{other} so that one could also investigate the relevance of
the detailed modeling of the coupling to the lattice, on the behaviour
of the carrier. Finally, addition of Anderson disorder is also
straightforward to implement in this approach, and would open a way to
investigate the competition between the localization  promoted by disorder and
the delocalizing effects of the e-ph coupling, away from
perturbational regimes. Indeed,  we believe that the method we have
proposed and developed here can be used to study efficiently yet quite
accurately a  varied range
of interesting problems.

\begin{acknowledgments} We thank Sarah Burke for 
suggesting this problem to us. This work
was supported by  
NSERC and QMI.
\end{acknowledgments}



\begin{thebibliography}{99}
\bibitem{comment} Note that the number of phonons at site $n_k$ is not
  necessarily $p+1$, since in principle it is possible that the
  electron has been at site $n_k$ before and created a phonon, after
  which it moved to create phonons at other sites and only then
  returned to $n_k$ to create a cloud. As already argued such
  processes are not very likely. To the extent that they happen, their
  effect is to replace $p\rightarrow p+1$ if one extra phonon was
  already there, etc. For large couplings where the typical number $p$
  of phonons in the cloud can be large, this will have relatively
  small numerical consequences.
\end{thebibliography}

\end{document}